\documentclass[%
superscriptaddress,
 preprint,
 amsmath,amssymb,
 aps,
 jcp,
 letterpaper,
]{revtex4-2}

\usepackage{graphicx}
\usepackage{dcolumn}
\usepackage{bm}

\usepackage{array}
\usepackage{braket}
\usepackage{droidsans}
\usepackage[T1]{fontenc}

\usepackage{epstopdf}
\usepackage{multirow}

\def\Ddosh{D$_{2\text{h}}$}
\def\Dseish{D$_{6\text{h}}$}
\def\Ag{A$_{\text{g}}$}

\def\Bunog{B$_{1\text{g}}$}

\def\Bdosu{B$_{2\text{u}}$}
\def\Btresu{B$_{3\text{u}}$}

\newcommand{\met}{\mathrm}
\newcommand{\mat}{\mathbf}

\usepackage[unicode]{hyperref}
\hypersetup{
   unicode=true,          
   plainpages=false,
   colorlinks=true,       
   citecolor=blue,        
}

\def\sinfo{\text{Supplementary Information}}

\begin{document}

\title{%
Coronene: a model for ultrafast dynamics in graphene nanoflakes and PAHs
}

\author{Alberto Mart\'in Santa Dar\'ia}
\author{Lola González-Sánchez}
\author{Sandra Gómez}
\email{sandra.gomez@usal.es}

\affiliation{
Departamento de Qu\'imica F\'isica, Universidad de Salamanca https://ror.org/02f40zc51, Spain \\}

\date{\today}
\begin{abstract}
  \noindent Assuming a delta pulse excitation, quantum wavepackets are propagated on the excited state manifold in the energy range from 3.4-5.0 eV for coronene and 2.4-3.5 eV for circumcoronene to study the time evolution of the states as well as their lifetimes. The full-dimensional (102 and 210 degrees of freedom for coronene and circumcoronene respectively) non-adiabatic dynamics simulated with the ML-MCTDH method on twelve coupled singlet electronic states show that the different absorption spectra are only due to electronic delocalisation effects that change the excited state energies, but the structural dynamics in both compounds are identical. Breathing and tilting motions drive the decay dynamics of the electronic states away from the Frank-Condon region independently of the size of the aromatic system. This promising result allows the use of coronene as a model system for the dynamics of larger polycyclic aromatic hydrocarbons (PAHs) and graphene one dimensional sheets or nanoflakes.
  \begin{figure}[h]
    \centering
    \includegraphics[width=0.6\linewidth]{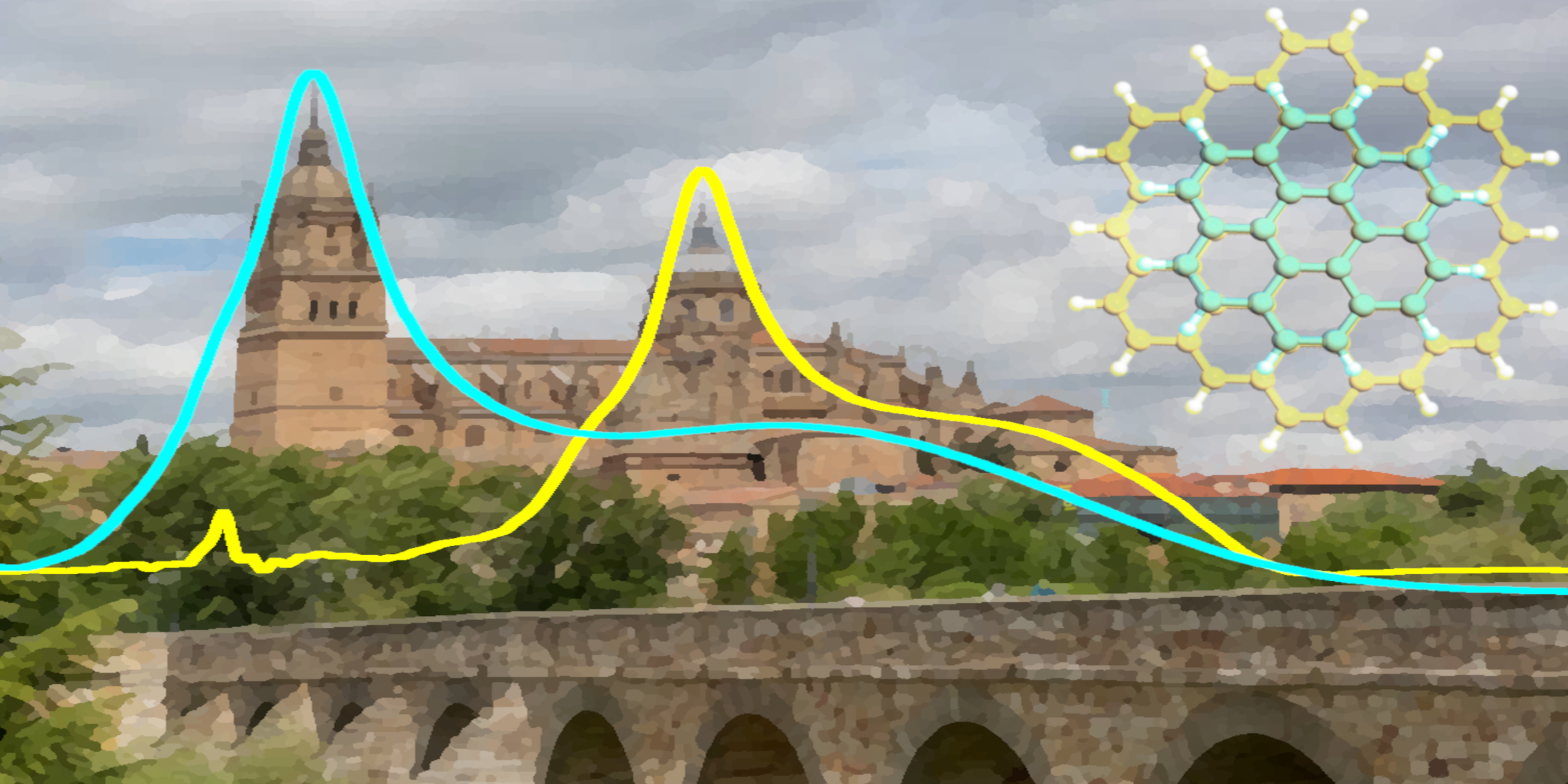}
    \label{fig:toc}
\end{figure}

\end{abstract}

\maketitle 

\clearpage

\section{Introduction}
\noindent


Photoinduced processes play a crucial role in many fields of chemistry, physics, biology, and material sciences, ranging from their involvement in vision to anticancer therapies and optoelectronics.\cite{Domcke2012,Lan2019,Ostroverkhova2016,Xu2018,Keller2021} 
Understanding such processes is essential to be able to design new compounds and tune the optical and electronic properties of materials, enabling a desired functionalisation. Particularly in the field of optoelectronic devices such as solar cells, light-emitting diodes (LEDs), and lasers, photoinduced processes are fundamental, influencing their performance and efficiency. 
To conveniently tune their performance, one must understand the behavior of the excited states of the target materials and study their time evolution. 
Studying these processes enhances our understanding of fundamental physical phenomena, facilitates the development of new theoretical models, and guides the design of novel materials with tailored properties.\cite{Pham2021,Bassan2021,Marin2023}

Polycyclic aromatic hydrocarbons (PAHs) are $\pi$-conjugated organic molecules that consist of multiple fused aromatic rings. 
They are naturally found as products from the incomplete combustion of organic materials, forest fires, and volcanic eruptions.\cite{giger1974polycyclic,cerniglia1993biodegradation} 
Leaving aside their threat to humans health,\cite{kim2013review} in the last years they have become highly relevant in industrial applications such as the production of dyes and pigments and pharmaceuticals~\cite{Huberman1976,Berenbeim2019} due to their photostability and photophysical properties.\cite{shao2020tuning,huisken2013} These compounds have also been found to be a key piece in the understanding the chemistry of the interstellar medium.\cite{Malloci2004}

Moving towards the solid state, graphene quantum dots (GQDs) are nanostructures composed of small fragments of graphene, which is a two-dimensional carbon allotrope, typically showing lateral dimensions ranging from a few to several hundred nanometers.\cite{bacon2014graphene} 
GQDs possess unique optical, electronic, and chemical characteristics, distinct from bulk graphene, presumably arising from their size, edge structures, and quantum confinement effects.\cite{shen2012graphene,zhu2012control,sun2013recent} 
In recent years, it has been proposed that their tunable photoluminescence may be related to the method of synthesis, attributed to doping effects, surface functional groups, and the presence of molecular fluorophores,\cite{jin2013tuning,yuan2019carbon,hola2014carbon,li2019carbon,kozak2016photoluminescent,liu2019carbon}
which has been widely investigated also by theoretical computations, recently reviewed by M. Langer \emph{et al.} in Ref.~\citenum{langer2021progress}.
As a consequence of their ashtonishing properties, GQDs have applications in optoelectronics,\cite{yuan2019carbon, arcudi2017porphyrin,tetsuka2016molecularly,semeniuk2019future} energy storage,\cite{liu2020graphene,li2015carbon} sensing,\cite{fan2015fluorescent,sun2013recent} and biomedicine.\cite{kim2018selective,kadian2021recent}

Graphene nanoflakes, also known as graphene nanoribbons or nanostrips, are narrow and elongated structures of graphene,
and similarly to GQDs, they can be synthesized with different widths, lengths, and edge structures, resulting in diverse electronic properties.\cite{kuc2010structural,mansilla2016optical} 
Their edge-state-dependent properties are due to quantum confinement along the width of the nanoflakes. 
In general, they are larger than GQDs, and therefore the quantum confinement effects may be less pronounced.

Coronene and circumcoronene are both examples of polycyclic aromatic hydrocarbons that are often used as models for studying the properties of larger PAHs, GQDs, and graphene nanoflakes.\cite{shi2019high}
Coronene is a PAH consisting of a planar molecule with a hexagonal shape, composed of seven fused benzene rings. It has a well-defined structure and is often used as a model system to study the properties of extended aromatic compounds due to its relative simplicity. Coronene is known for its fluorescence properties and has been extensively studied for its electronic, optical, and chemical properties.\cite{Bermudez1986} It is also used as a benchmark molecule in theoretical calculations and experimental studies.\cite{Pat1942, HiSaAr14,Timofeeva21,MoDaFr03} 
UV-VIS results show a band at 4.1 eV, with a shoulder displaced to the blue (4.3 eV). A forbidden band, allowed only by symmetry breaking, appears at 3.6 eV. 

Circumcoronene is a larger PAH composed of a circular arrangement of benzene rings. It is formed by the fusion of multiple coronene units, resulting in a larger and more complex structure. It contains exactly double the amount of carbon and hydrogen atoms than coronene (48 carbons and 24 hydrogens). Due to its size and extended $\pi$  conjugation, circumcoronene exhibits unique electronic and optical properties. It is often employed as a model system to investigate the properties of graphene nanoflakes because of its structural similarity (and its larger size compared to coronene) and the possibility of tuning its electronic structure through chemical modifications. The synthesis of circumcoronene has always remained a challenge and it has been recently synthesized in solution,\cite{Zhou2023} although a UV-VIS spectrum is not yet available.

Both coronene and circumcoronene have been widely studied in various fields, including materials science, organic electronics, and nanotechnology, as they provide valuable insights into the behavior of PAHs and graphene nanoflakes.
Due to their reduced size compared to the material structures of interest, 
 one can use more accurate theoretical methods to deeply understand their properties. Therefore, they serve as important model systems for understanding the fundamental properties and potential applications of larger carbon-based materials.\cite{shi2019high}
Although there are excited state dynamical studies on smaller PAHs, such as benzene, \cite{ben09,ben17} naphthalene dimer, \cite{naph22} anthrazene \cite{ant15} and pyrene,\cite{py22} the field is lacking an exhaustive work on the excited state behavior of coronene and circumcoronene. 
The purpose of this article is to compare the excited state dynamics on coronene and circumcoronene molecules, to assess the feasibility of using the coronene molecule as a model system for larger polycyclic aromatic hydrocarbons. 
\section{Theoretical and computational framework}

\subsection{Electronic structure details \label{sec:compdetails}}
\noindent
To compute vertical energies, gradients and couplings needed throughout this work, we benchmarked several TD-DFT functionals and basis sets against the wave function method EOM-CCSD/cc-pVDZ for the coronene molecule. For the CCSD ground state reference, the symmetry was chosen to be \Ddosh, the highest symmetry available. 
The Q-Chem package \cite{qchem2015} was used to compute the reference EOM-CCSD vertical energies, whereas
and Gaussian 16 \cite{g16} was chosen to compute the vibrational frequencies of both systems. For every other calculation, i.e., Wigner-displaced geometries, parametrisation of the vibronic coupling Hamiltonian and on-the-fly dynamics, we used the ORCA 5.0 electronic structure program.\cite{Orca12,Orca22} 
The details about the benchmark we performed are included in the \sinfo, and they are supported by a previous work by B. Shi \emph{et al.}\cite{shi2019high} 
The TD-B3LYP-D3/def2-SVP method was proven to provide an excellent agreement with respect to experimental results while keeping the computational time
feasible. 
For circumcoronene, we used the same combination to compute the parametrised potentials: TD-B3LYP-D3/def2-SVP.

\subsection{The vibronic coupling model \label{sec:lvc}}
\noindent
To describe the potential energy surfaces of the systems, we constructed an approximate Hamiltonian as a first order Taylor expansion of the vertical electronic state energies calculated at the Frank-Condon (FC) point.  

A diabatic basis ensures a smooth description of the potential energy surfaces and mass- 
and frequency-weighed normal mode coordinates $Q_\alpha (\alpha = 1,\ldots,f)$, where $f$ is the number of vibrational degrees of freedom (DOFs) of the system, are chosen as the coordinate basis to expand the potential energy surfaces. Tuning modes carry the gradient, driving the molecules away from the FC region, whereas coupling modes mix pairs of electronic states and are responsible for the population transfer.
When this expansion is truncated in the first order, the method is known as the linear vibronic coupling (LVC) model, and the Hamiltonian looks as follows:
\begin{equation}
    \label{eq:lvc}
    \pmb{H} = \pmb{H}^{(0)} + \pmb{W}^{(0)} + \pmb{W}^{(1)} \; .
\end{equation}

The zeroth-order Hamiltonian $\pmb{H}^{(0)}$ is defined by the diagonal kinetic energy operator of the $N$ electronic states to be included,
\begin{equation}
    \label{eq:lvc_h0}
    H^{(0)}_{ij} (\pmb{Q}) = T_N \delta_{ij} = 
    - \frac{1}{2} \sum_\alpha \omega_\alpha  \frac{\partial^2}{\partial Q_\alpha^2} \; .
\end{equation}

A simple way of defining the zeroth-order potential matrix $\pmb{W}^{(0)}$ is to use the harmonic approximation shifted by the vertical excitation energies $E_i$ at the Frank-Condon (FC) geometry 
\begin{equation}
    \label{eq:lvc_w0}
    W^{(0)}_{ij} (\pmb{Q}) = \delta_{ij} (E_i + V^{\text{HO}} )  = 
    \delta_{ij} \bigg( E_i + \frac{1}{2} \sum_\alpha \omega_\alpha Q_\alpha^2 \bigg) \; ,
\end{equation}
while $\pmb{W}^{(1)}$, the first-order corrections of the potential matrix are defined using the following expansion:
\begin{equation}
    \label{eq:lvc_w1}
    W^{(1)}_{ij} (\pmb{Q}) = \delta_{ij} \sum_\alpha \kappa_\alpha^{(i)} Q_\alpha 
    + (1-\delta_{ij}) \sum_\alpha \lambda_\alpha^{(ij)} Q_\alpha \; .
\end{equation}
The $\kappa_\alpha^{(i)}$ linear intra-state vibronic couplings  carry the gradients (along the $\alpha$ tuning modes) of the $i$ excited potential energy surface at the FC point. 
The $\lambda_\alpha^{(ij)}$ linear inter-state vibronic constants measure the coupling between each pair of electronic states $i$ and $j$ along the normal mode $\alpha$.

In this work, we used the LVC model as implemented in the SHARC molecular dynamics package \cite{sharc11,sharc15,sharc18}, which generates all the LVC model parameters\cite{sharc19} from 
(a) the calculation of the ground-state frequencies at the gas-phase optimised geometries of coronene and circumcoronene and 
(b) $2(3N-6)+1$ excited state energy calculations of displaced geometries in positive and negative directions along every vibrational mode from the equilibrium configuration. From these computations, diabatic state gradients and couplings were computed from numerical derivatives, as detailed in Ref. \citenum{sharc19}. As we are dealing with systems of non-abelian symmetry groups, a rotation of the Hamiltonian was applied to the Hamiltonian to correct spurious couplings, as described in Ref. \citenum{gomez2019}.

\subsection{The ML-MCTDH wavepacket propagation method}
The nuclear dynamics of the system can be studied by solving the time-dependent Schrödinger equation (TDSE),
\begin{equation}
    \label{eq:tdse}
    i \frac{\partial}{\partial t} \Psi(\pmb{Q},t) = H \Psi(\pmb{Q},t)
\end{equation}
where the Hamiltonian operator $H$ can be generalized to describe a system characterized by a set of coupled potential energy surfaces in the diabatic picture, like in Eq.~\ref{eq:lvc}.
The total wavefunction of the system is a function of the normal modes and its definition depends on the method we use to solve this first-order differential equation.

A common approach to solve the TDSE is the multiconfigurational time dependent Hartree (MCTDH) method,\cite{bec00:1,mey09} which expands the total wave function in a set of time-dependent basis functions  $\psi_{p}^{(\alpha)}$ as
\begin{equation}
    \label{eq:mctdh}
    \Psi(\pmb{Q},t) = \sum_{p_1} \cdots \sum_{p_f} 
    A_{jp_1 \ldots p_f}(t) \cdot  \psi_{p_1}^{(1)} (Q_1,t) \cdots  \psi_{p_f}^{(f)} (Q_f,t)
\end{equation}
which are known as single-particle functions (SPFs) and are expanded in a time-independent primitive basis set $\chi^{(\alpha)}$, constructed using a DVR\cite{lig85:1400} grid,
\begin{equation}
    \label{eq:spfs}
    \psi_{p}^{(\alpha)} (Q_\alpha,t) =
    \sum_q a_{pq}(t) \cdot \chi_q^{(\alpha)}(Q_\alpha) \; .
\end{equation}

Due to the exponential scaling of the computational effort with respect to the dimensionality of the system, MCTDH is normally used for systems with no more than 30 vibrational degrees of freedom. 
For higher-dimensional systems, the multi-layer MCTDH (ML-MCTDH) \cite{wan03:1289} approach can be used, where a recursive layering structure is constructed in a way that the SPFs are expanded in a set of time-dependent SPFs and only the last layer is expanded using the time-independent DVR primitive basis, still ensuring perfect convergence to the exact quantum dynamics result at the nuclear basis limit. 

All these methodologies are the core of the Quantics package \cite{WoGiRi15,Wo20}, used throughout this work to study the quantum dynamical properties of coronene and circumcoronene.

\subsection{Trajectory Surface Hopping}
The Tully surface hopping (TSH)\cite{Tully1971, Tully1990} method propagates the nuclei following the Newton equations of motion
\begin{equation}
  M_A\frac{\met{d}^2}{\met{d}t^2}\mat{R}_A(t)=-\frac{\met{d} V_{\alpha}(\mat{R}_A)}{\met{d} \mat{R}}\left(\mat{R}(t)\right)\label{eq:theo:newton}
\end{equation}
where the atom $A$ is moving in a single potential energy surface ($V_i\left(\mat{R}(t)\right)$) pushed by forces corresponding to its gradient. Compared to MCTDH, instead of a full nuclear wavepacket, TSH consists of individual trajectories that move completely uncoupled from the bundle.

The propagation from one-time step to another is performed using the velocity Verlet algorithm,\cite{Verlet1967} where velocities $\mat{v}_A(t)$ and the atomic coordinates $\mat{R}_A(t)$ change according to
\begin{align}
  \mat{R}_A(t+\Delta t) &= \mat{R}_A(t) + \mat{v}_A(t)\Delta t + \frac{1}{2M_A}\frac{\met{d} V_{\alpha}(\mat{R}_A)}{\met{d} \mat{R}}(t)\Delta t^2, \\
  \mat{v}_A(t+\Delta t) &= \mat{v}_A(t) + \frac{1}{2M_A}\frac{\met{d} V_{\alpha}(\mat{R}_A)}{\met{d} \mat{R}}(t)\Delta t+\frac{1}{2M_A}\frac{\met{d} V_{\alpha}(\mat{R}_A)}{\met{d} \mat{R}}(t+\Delta t)\Delta t.
\end{align}

The nuclei are assumed to move adiabatically most of the time, and in some particular regions where the coupling with other electronic states is high, undergo non-adiabatic transitions known as hops, thus the name surface hopping.\cite{Tully1971, Tully1990} At every time step, the nuclei have the choice to change the electronic state based on a probability to hop that depends on the coefficients of the electronic states involved. The electronic wavefunction is therefore expressed as a sum of different electronic states with time-dependent coefficients using the Born-Huang expansion. Inserting this wavefunction ansatz in the TDSE yields to the following equations for the propagation of the coefficients:

\begin{equation}
  \frac{\met{d}c_{\beta}(t)}{\met{d}t} = -\sum_{\alpha}\Bigg[\met{i}\underbrace{\Braket{\Psi_{\beta}^{\mathrm{el}}|\hat{H}^{\met{el}}|\Psi_{\alpha}^{\mathrm{el}}}}_{ H_{\beta\alpha}}+\underbrace{\Braket{\Psi_{\beta}^{\mathrm{el}}|\frac{\met{d}}{\met{d}t}|\Psi_{\alpha}^{\mathrm{el}}}}_{ K_{{\beta}\alpha}} \Bigg]c_{\alpha}(t) \; , \label{eq:theo:coeffs}
\end{equation}

\noindent where the Hamiltonian term is directly calculated from the precomputed PES or via electronic structure methods, and the second term is calculated from the NAC between electronic states and velocities: $K_{\beta\alpha}= \Braket{\Psi_{\beta}^{\mathrm{el}}|\nabla_R|\Psi_{\alpha}^{\mathrm{el}}}\mat{v}_R$.
Throughout this work, we have used the SHARC implementation of TSH\cite{Mai2015b} both calculating electronic quantities on-the-fly and on precomputed potentials, as well as the more accurate wavepacket method ML-MCTDH on the same potentials. The reason for using the less-accurate method TSH was twofold: first, to assess the role of triplets since the spin orbit coupling was negligible at the Frank-Condon geometry and, second, to compare whether dynamics on precomputed potentials were a good approximation to the dynamics computed on-the-fly. 
The TSH calculations used the ``fewest switches'' algorithm,\cite{Tully1990} along with the decoherence correction of Granucci and Persico.\cite{Granucci2010} The time-step was chosen to be 0.5 fs. The hopping probabilities were obtained from state overlaps.\cite{Plasser2016} 400 initial structures and velocities were generated from a Wigner distribution.
More system-specific computational details are provided at the begining of the corresponding sections.
%


\section{Results: Quantum dynamics of coronene \label{sec:coro}}

\subsection{Vertical absorption and static results} \label{sec:coro_abinitio}
\noindent

Using the level of theory introduced in Sec.~\ref{sec:compdetails}, 
we computed the corresponding vertical excitations from the singlet ground state (X, S$_0$) of coronene, collected in Table~\ref{tab:vert-excited} in energetic order following the TD-B3LYP-D3 results.
Besides the experimental reference, \cite{HiSaAr14} 
Table~\ref{tab:vert-excited} includes also the electronic character of every excited state, providing the molecular orbitals (MOs) with the largest contributions.
\begin{table}
    \caption{%
      Vertical excitation energies, in eV, and their corresponding oscillator strengths obtained using EOM-CCSD/cc-pVDZ 
      level of theory and TD-B3LYP-D3/def2-SVP for coronene in \Ddosh\ symmetry. 
      The experimental absorption maxima are taken from Ref.~\citenum{HiSaAr14}.
      \label{tab:vert-excited}
    }
    \centering
    \small
    \begin{tabular}{@{} c@{\ } c@{ } c@{\ \ } c@{\ \ } l@{\ \ } l@{\ } l@{\ \ } l@{\ \ } c@{\ } c@{\ \ } l@{\ \ } @{}}
      \hline\hline\\[-0.65cm]
        State$^\text{a}$   &    Symm.$^\text{b}$ & Character$^\text{c}$ &
        \multicolumn{2}{l}{TD-B3LYP-D3/def2-SVP} 
        & & \multicolumn{2}{l}{EOM-CCSD/cc-pVDZ}  
        & & \multicolumn{2}{l}{Expt.\cite{HiSaAr14}} \\
      \hline\\[-1.1cm]
\multirow{2}{0.5cm}{S$_0$} & \multirow{2}{0.5cm}{X(A$_{\text{g}})   $} & & ~~~\multirow{2}{0.5cm}{0.00} &       \multirow{2}{0.5cm}{}     &&      ~~~\multirow{2}{0.5cm}{0.00}    &       \multirow{2}{0.5cm}{}     &&  \multirow{2}{0.5cm}{    }     &       \multirow{2}{0.5cm}{  }        \\[-0.45cm]
                                         &                                       &       &   \\[-0.1cm]
\multirow{2}{0.5cm}{S$_1$}        &     \multirow{2}{0.5cm}{1\Bdosu$\biggl\{$}   &      4b$_{1\text{u}} \rightarrow$ 4b$_{3\text{g}}$           &      ~~~\multirow{2}{0.5cm}{3.22}    &       \multirow{2}{0.5cm}{(0.00)}     &&      ~~~\multirow{2}{0.5cm}{4.06}    &       \multirow{2}{0.5cm}{(0.00)}     &&  \multirow{2}{0.5cm}{    }     &       \multirow{2}{0.5cm}{  }        \\
                                &                                       &       2a$_{\text{u}} \rightarrow$ 4b$_{2\text{g}}$   \\[-0.1cm]
\multirow{2}{0.5cm}{S$_2$}        &     \multirow{2}{0.5cm}{1\Btresu$\biggl\{$}    &    4b$_{1\text{u}} \rightarrow$ 4b$_{2\text{g}}$            &      ~~~\multirow{2}{0.5cm}{3.51}    &       \multirow{2}{0.5cm}{(0.00)}     &&      ~~~\multirow{2}{0.5cm}{3.30}    &       \multirow{2}{0.5cm}{(0.00)}     &&  \multirow{2}{0.5cm}{3.65}   &       \multirow{2}{0.5cm}{(0.51)}        \\
                                &                                       &       2a$_{\text{u}} \rightarrow$ 4b$_{3\text{g}}$   \\[-0.1cm]
\multirow{2}{0.5cm}{S$_3$}        &     \multirow{2}{0.5cm}{1\Bunog$\biggl\{$}    &     4b$_{1\text{u}} \rightarrow$ 3a$_{\text{u}}$            &      ~~~\multirow{2}{0.5cm}{4.32}    &       \multirow{2}{0.5cm}{(0.00)}     &&      ~~~\multirow{2}{0.5cm}{4.55}    &       \multirow{2}{0.5cm}{(0.00)}     &&  \multirow{2}{0.5cm}{    }   &       \multirow{2}{0.5cm}{    }       \\
                                &                                       &       2a$_{\text{u}} \rightarrow$ 5b$_{1\text{u}}$   \\[-0.1cm]
\multirow{2}{0.5cm}{S$_4$}        &     \multirow{2}{0.5cm}{~~1\Ag $\biggl\{$}    &     2a$_{\text{u}} \rightarrow$ 3a$_{\text{u}}$            &      ~~~\multirow{2}{0.5cm}{4.32}    &       \multirow{2}{0.5cm}{(0.00)}     &&      ~~~\multirow{2}{0.5cm}{4.55}    &       \multirow{2}{0.5cm}{(0.00)}     &&  \multirow{2}{0.5cm}{    }   &       \multirow{2}{0.5cm}{    }       \\
                                &                                       &       4b$_{1\text{u}} \rightarrow$ 5b$_{1\text{u}}$   \\[-0.1cm]
\multirow{2}{0.5cm}{S$_5$}        &     \multirow{2}{0.5cm}{~~2\Ag$\biggl\{$}    &      2a$_{\text{u}} \rightarrow$ 3a$_{\text{u}}$            &      ~~~\multirow{2}{0.5cm}{4.35}    &       \multirow{2}{0.5cm}{(0.00)}     &&      ~~~\multirow{2}{0.5cm}{4.82}    &       \multirow{2}{0.5cm}{(0.00)}     &&  \multirow{2}{0.5cm}{    }   &       \multirow{2}{0.5cm}{    }       \\
                                &                                       &       4b$_{1\text{u}} \rightarrow$ 5b$_{1\text{u}}$   \\[-0.1cm]
\multirow{2}{0.5cm}{S$_6$}        &     \multirow{2}{0.5cm}{2\Bdosu$\biggl\{$}   &      4b$_{1\text{u}} \rightarrow$ 4b$_{3\text{g}}$            &      ~~~\multirow{2}{0.5cm}{4.38}    &       \multirow{2}{0.5cm}{(0.85)}     &&      ~~~\multirow{2}{0.5cm}{4.85}    &       \multirow{2}{0.5cm}{(1.14)}     &&  \multirow{2}{0.5cm}{4.09}   &       \multirow{2}{0.5cm}{(2.24)}     \\
                                &                                       &       2a$_{\text{u}} \rightarrow$ 4b$_{2\text{g}}$    \\[-0.1cm]
\multirow{2}{0.5cm}{S$_7$}        &     \multirow{2}{0.5cm}{2\Btresu$\biggl\{$}    &    4b$_{1\text{u}} \rightarrow$ 4b$_{2\text{g}}$            &      ~~~\multirow{2}{0.5cm}{4.38}    &       \multirow{2}{0.5cm}{(0.85)}     &&      ~~~\multirow{2}{0.5cm}{4.85}    &       \multirow{2}{0.5cm}{(1.14)}     &&  \multirow{2}{0.5cm}{4.09}   &       \multirow{2}{0.5cm}{(2.24)}     \\
                                &                                       &       2a$_{\text{u}} \rightarrow$ 4b$_{3\text{g}}$    \\[-0.1cm]
\multirow{2}{0.5cm}{S$_8$}        &     \multirow{2}{0.5cm}{2\Bunog$\biggl\{$}    &     4b$_{1\text{u}} \rightarrow$ 3a$_{\text{u}}$            &      ~~~\multirow{2}{0.5cm}{4.42}    &       \multirow{2}{0.5cm}{(0.00)}     &&      ~~~\multirow{2}{0.5cm}{4.89}    &       \multirow{2}{0.5cm}{(0.00)}     &&  \multirow{2}{0.5cm}{    }   &       \multirow{2}{0.5cm}{    }\\
                                &                                       &       2a$_{\text{u}} \rightarrow$ 5b$_{1\text{g}}$   \\[-0.1cm]
\multirow{2}{0.5cm}{S$_9$}        &     \multirow{2}{0.5cm}{3\Bunog$\biggl\{$}    &     3b$_{3\text{g}} \rightarrow$ 4b$_{2\text{g}}$            &      ~~~\multirow{2}{0.5cm}{4.58}    &       \multirow{2}{0.5cm}{(0.00)}     &&      ~~~\multirow{2}{0.5cm}{5.27}    &       \multirow{2}{0.5cm}{(0.00)}     &&  \multirow{2}{0.5cm}{    }   &       \multirow{2}{0.5cm}{    }\\
                                &                                       &       3b$_{2\text{g}} \rightarrow$ 4b$_{3\text{g}}$    \\[-0.1cm]
\multirow{2}{0.5cm}{S$_{10}$}     &     \multirow{2}{0.5cm}{4\Bunog $\biggl\{$}    &    3b$_{3\text{g}} \rightarrow$ 4b$_{2\text{g}}$             &      ~~~\multirow{2}{0.5cm}{4.70}    &       \multirow{2}{0.5cm}{(0.00)}     &&      ~~~\multirow{2}{0.5cm}{5.34}    &       \multirow{2}{0.5cm}{(0.00)}     &&  \multirow{2}{0.5cm}{    }   &       \multirow{2}{0.5cm}{    }\\
                                &                                       &       3b$_{2\text{g}} \rightarrow$ 4b$_{3\text{g}}$   \\[-0.1cm]
\multirow{2}{0.5cm}{S$_{11}$}     &     \multirow{2}{0.5cm}{~~3\Ag $\biggl\{$}    &     3b$_{3\text{g}} \rightarrow$ 4b$_{3\text{g}}$             &      ~~~\multirow{2}{0.5cm}{4.70}    &       \multirow{2}{0.5cm}{(0.00)}     &&      ~~~\multirow{2}{0.5cm}{5.27}    &       \multirow{2}{0.5cm}{(0.00)}      &&  \multirow{2}{0.5cm}{   }   &       \multirow{2}{0.5cm}{    }\\
                                &                                       &       3b$_{2\text{g}} \rightarrow$ 4b$_{2\text{g}}$   \\
      \hline\hline
    \end{tabular}
\begin{flushleft}
$^\text{a}$~Adiabatic state labeling in energetic order at the FC region based on the TD-B3LYP-D3 results. \\
$^\text{b}$~Diabatic symmetry labels within the \Ddosh\ molecular symmetry group of the singlet excited states (at the TD-B3LYP-D3/def2-SVP level). 
\\
$^\text{c}$~The transitions between MOs of coronene that contribute the most ($> 90\%$) to each excited state. The symmetry labels correspond to the \Ddosh\ molecular symmetry group.
\end{flushleft}
\end{table}
The graphical representations of these MOs are included in Fig.~\ref{fig:circum_mo}, where the last 4 occupied orbitals are in the bottom figure, and the first 4 virtual orbitals are in the top row. 
The MOs represented in Fig.~\ref{fig:circum_mo} are degenerate in the \Ddosh\ symmetry group in the following pairs of MOs: (LUMO$+$3, LUMO$+$2), (LUMO$+$1, LUMO), (HOMO, HOMO$-$1), (HOMO$-$2, HOMO$-$3).

Among all the vertical transitions involving these orbitals, only two of them show oscillator strength at 4.4 eV, which corresponds to the excitation to the 2\Bdosu (S$_6$) and 2B$_{3\text{u}}$ (S$_7$) singlet excited states, matching the maximum absorption peak of the experimental spectrum around 4 eV. These two bright states are degenerate in \Ddosh, and they correspond to a E$_{1\rm u}$ state in \Dseish.\cite{shi2019high}
There is another absorption peak in the experimental spectrum at 3.6 eV which appears dark in the FC region and corresponds to the excitation to the 1\Btresu\ (S$_2$) state. 
The transition to this state is forbidden by symmetry, but it becomes accessible through the vibronic couplings between states. 
This effect has been observed by generating a set of initial conditions displaced from the equilibrium configuration using a Wigner distribution based on the S$_0$ state frequencies computed at the TD-B3LYP-D3/def2-SVP level of theory. 

\newpage

\begin{figure}
  \begin{center}
    \includegraphics[width=0.7\linewidth]{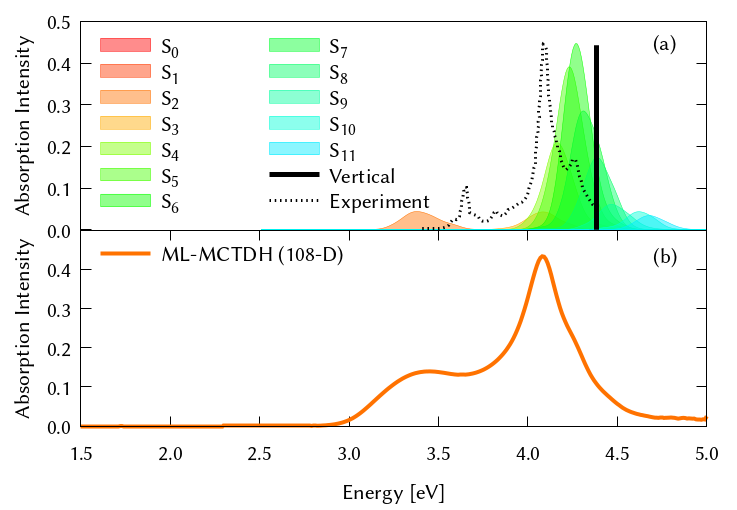}
    \end{center}     
    \caption{%
      Computed absorption spectrum of coronene. 
      Top panel (a) shows the static spectra computed by vertical excitations from the FC region using TD-DFT (black line)
      and including vibronic coupling by computing vertical excitations  
      from 200 initial conditions selected from a Wigner distribution. The experimental absorption spectrum is taken from Ref.~\citenum{HiSaAr14}.
      Lower panel (b) represents the dynamic absorption spectra obtained from the Fourier Transformation (FT) $\langle\Psi(t)|\Psi(t=0)\rangle$ of the autocorrelation function of the quantum dynamics starting from 1\Btresu\ (S$_2$) excited state computed with ML-MCTDH and including the 102 normal modes of coronene. }
    \label{fig:coro_spec}
\end{figure}

The vibrationally-resolved absorption spectrum obtained by exciting from all the initial configurations is represented in the upper panel (a) of Fig.~\ref{fig:coro_spec}, where a new bright band corresponding to excitation to the 1\Btresu\ (S$_2$) state is observed around 3.4 eV.
Only singlet states are included in this work due to the negligible Spin-Orbit Couplings (SOCs) present at the equilibrium configuration. The initial parametrisation of the Hamiltonian included also twenty triplet states, i.e., sixty electronic states more. Every matrix element was very small, indicating that the triplet states will not be populated at this geometry. To further analyse whether at different geometries changes that statement,  fifty four TSH on-the-fly trajectories were propagated for 20 fs, in which no population transfer to any of the triplet excited states is observed, at least at the short time scales analysed in this study.

After the static analysis of the excited states in coronene, an LVC Hamiltonian, as described in Sec.~\ref{sec:lvc}, has been constructed in order to approximate the analytical potential energy surfaces including twelve coupled singlet states projected onto the 102 nuclear degrees of freedom of the molecule.  

\subsection{Excited state quantum dynamics from the first absorption band: 1\Btresu\ (S$_2$) \label{sec:coro_s2}}
To study the dynamical evolution of the diabatic state manifold, excited state dynamics have been carried out with different methods from both bright excited states, S$_2$ and S$_6$/S$_7$, as well as from the dark states to show how well it compares to non-adiabatic dynamics in the circumcoronene system.

For the case of the TSH, the 54 on-the-fly trajectories were initialised based on oscillator strengths using an energetic window of 3.0 to 3.75 eV to excite to the first "forbidden" band and performed on four coupled singlet electronic states (S$_0$-S$_4$) and six triplets (T$_1$-T$_6$) to minimise the computational effort needed. 
The TSH dynamics on LVC potentials started from the same geometries as the on-the-fly dynamics and included 54 trajectories on twelve singlet and eleven triplet electronic states. 
The ML-MCTDH dynamics were initialised as the hartree product of the vibrational eigenstate of every degree of freedom. The dynamics on the twelve coupled electronic states was propagated using Runge-Kutta 5 for the nuclear degrees of freedom and short iterative Lanczos (SIL) for the electronic one. 4.6$\times 10^7$ primitive basis functions were used for the grid and the calculations included 1.2$\times 10^{35}$ configurations.  For the 8D reduced MCTDH dynamics the Bulirsch-Stoer extrapolation scheme  was used for the propagation of the SPFs and the SIL integrator for the A-coefficients and the electronic propagation. The total size of the grid for the converged results in this system was 3.25$\times 10^9$ and used 789 nuclear configurations.

\begin{figure}
  \begin{center}
    \includegraphics[width=0.7\linewidth]{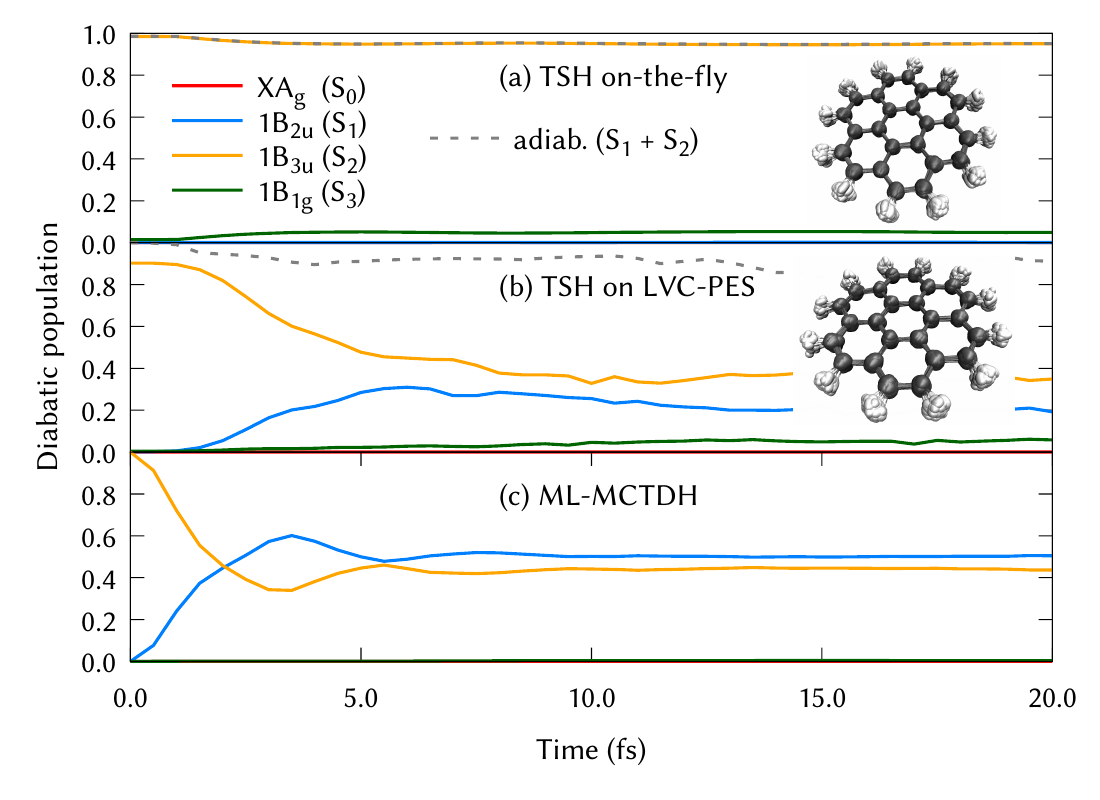}
    \end{center}     
    \caption{%
      Diabatic population transfer from the 1\Btresu\ (S$_2$) excited state of coronene using three different propagations:
      (a) On-the-fly Tully Surface Hopping classical trajectories (TSH). The diabatisation has been made just tracing back to the state character of the initial populated state and populations are collected as the square of the quantum amplitudes of the states, 
      (b) TSH classical trajectories using the LVC model for the PES and, 
      (c) ML-MCTDH using also the LVC model potentials.
      Insets of molecular structures of coronene visited during both the on-the-fly and on LVC potentials using the TSH method are included in the figures (a) and (b) with 8892 and 2213 overlapped structures respectively.
      }
    \label{fig:3pops}
\end{figure}

The experimental gas phase absorption spectrum of coronene was recorded by S. Hirayama \emph{et al.},~\cite{HiSaAr14} where the first absorption band of coronene was reported around $\sim$3.6 eV, which corresponds to excite the system to the second singlet excited state, 1\Btresu\ (S$_2$).
The population transfer from this excited state is shown in Fig.~\ref{fig:3pops} using the three different methods for propagating the nuclear dynamics: (a) on-the-fly TSH classical trajectories, (b) TSH classical trajectories on the LVC potentials and (c) ML-MCTDH quantum dynamics on the same LVC potentials. 

\newpage
In Fig.~\ref{fig:3pops}, the main difference between panels (a) and (b) is the PESs on which the classical nuclear trajectories were propagated. The more similar those panels, the better the LVC approximation in the PES construction. It is important to note that diabatic populations in panel (a) are not truly diabatic, but that the states have been labelled according to the diabatic state distribution that was present at the initial geometries, in a diabatic by ansatz manner. 
To clarify this fact, the combination of the S$_1$ and S$_2$ adiabatic populations is included in both (a) and (b) panels of Fig.~\ref{fig:3pops}, were a good agreement can be found.
TSH in this case does not give the same diabatic results on-the-fly or on precomputed potentials and both panels are substantially different. This may be surprising since the LVC model has been proven to be a good approximation in rigid systems with impeded torsions, as it is the case of the coronene molecule.
The performance of the LVC approximation has been also tested by inspecting the molecular configurations visited by the nuclear dynamics, represented as overimposed xyz structures in Fig.~\ref{fig:3pops} right hand side. 
This figure shows a fairly harmonic behavior of the coronene geometry during the classical trajectories of the nuclei both over the LVC potential and the `real' ab-initio potential.
Another proof-of-principle of the LVC approximation in this system is the fact that the absorption spectrum obtained with the PES is in good agreement with respect to the experimental, represented in the lower panel (b) of Fig.~\ref{fig:coro_spec}.
As we can see in Table.~\ref{tab:vert-excited}, the MOs involved in S$_1$ and S$_2$ are the same (with different relative weights), and they are degenerated in \Ddosh. To unravel why the diabatic population transfer differs, we checked the orbital character at the geometries visited along the on-the-fly dynamics and concluded that the differences arise due to a  trivial rotation of the degenerated MOs. 

In Fig.~\ref{fig:3pops}, the difference between panels (b) and (c) lies in the quantum treatment of the atomic nuclei in MCTDH in contrast to classical propagations during the TSH/LVC approach. From the comparison of these two panels, we can conclude that the TSH/LVC approach is a good approximation that covers qualitatively the population changes. However, the fine print of the dynamics of coronene is only provided by the most accurate method ML-MCTDH. 

For the ML-MCTDH method, all 102 vibrational modes of the system have been included. 
The ML-tree structure, available on the \sinfo,  has been optimized by looking at the largest frequency-weighted $\kappa_\alpha^{(i)}$ gradients and $\lambda_\alpha^{(ij)}$ couplings among states which involved the initial excited state 1\Btresu\ (S$_2$).\cite{LeGo20}

By using the same arguments, we can analyze those vibrations (normal modes) that contribute the most to the structure relaxation and drive the nuclear dynamics. 

The identification of the leading vibrations in the dynamics is crucial, not only for unraveling the nuclear dynamics of coronene but for validating this methodology for future work with larger systems like graphene nanoflakes, and to understand the absorption and emission properties of these systems in general.
For the case of coronene, the decay from S$_2$ is mostly due to the totally symmetric breathing mode $\nu_{24}$, which shows a negligible diabatic coupling between S$_2$ and S$_1$ but large gradient on the S$_2$ state. Another leading mode in this dynamics is $\nu_{19}$, the scissoring mode exhibiting the largest diabatic coupling. Two cuts of the diabatic PES along these two modes, along with their corresponding diabatic couplings are represented in the top panels of Fig.~\ref{fig:pes_cuts}.
The full-dimensional (full-D) dynamics have been compared with several reduced-dimensionality models, where only the most important modes have been included in a MCTDH calculation. 
We found the most relevant 8 modes which are the main responsible for the spectral features of the main band and during the first 10 fs of the dynamics. We systematically added modes to this 8D model trying to find a larger system to recover the full-D dynamics up to 50 fs. This was not possible, since there are many modes which contribute with small couplings to the Hamiltonian and cannot be left out. The spectrum resulting from the 8D dynamics and state populations compared to the full D results are depicted in the SI. The eight modes included are the three main breathing modes (full breathing v24, inner benzene breathings v60 and v77), four tilting modes (v19, v64, v79 and v88) and a CH stretching mode (v108). Note that the numeration of the modes started at v7 after removing the six translations and rotations.

After the vertical excitation from the ground state (GS) to the 1\Btresu\ (S$_2$) state, the diabatic population transfer from this excited state is represented in Fig.~\ref{fig:allpops} (top left panel) to up to 50 fs, 
where it can be seen that the population is equally distributed between S$_2$ and S$_1$ states already after the first 15 fs.
%
When classical trajectories are used for the nuclei motion, a population transfer is also observed from the S$_2$ to S$_1$ state. However, the time scales observed in panels b) and c) of Fig.~\ref{fig:3pops} are slightly different. Slower time scales are often observed when running classical trajectories, since they need to move towards the intersection, whereas a wavepacket has grid functions i.e., the possibility of population, in every part of the conformational space.
From this comparison, it is suggested that the quantum effects play an important role in the nuclear motion over the PESs. We could also  extract that electronic populations may not be the best description when dealing with high symmetry molecules, since arbitrary rotations impede a correct assignment of state characters.

\subsection{Excited state quantum dynamics from every electronic state \label{sec:coro_dyn}}

To complete the dynamical study while considering the results obtained from the analysis of the quantum dynamics from the 1\Btresu\ (S$_2$), 
ML-MCTDH has been carried out to unravel the quantum dynamics from every electronic excited state of coronene 
in the relevant energy range (below 5 eV). 
By exciting the system to each of the first 8 singlet states, we followed the corresponding diabatic populations.
The corresponding figures of these population transfers are collected in the \sinfo\ (Figs. S9-S16).

Since the excited states 2\Btresu\ (S$_6$) and 2\Bdosu\ (S$_7$) are responsible of the most intensive peak 
in the experimental absorption spectrum, the quantum dynamics from these states are especially relevant. 
Due to the symmetry reduction in this calculations, 2\Btresu\ and 2\Bdosu, which are degenerate in \Ddosh, 
correspond to the same state in the real symmetry group \Dseish. 
This is confirmed by the fact that the quantum dynamics from both excited states are almost identical. 
After the excitation to these states, $\sim 70\%$ of the population is transferred during the first 50 fs. 
From 2\Btresu\ (S$_6$), the decay seems slightly faster as compared to starting the dynamics on the 2\Bdosu\ (S$_7$).
Similar behavior is found for the diabatic population transfer from 1\Bunog\ (S$_3$) and 1\Ag\ (S$_4$) states, which are also degenerated in \Ddosh. 
The likeness between them is stronger during the first 10 fs, going from S$_3$ and S$_4$ to the degenerated pair S$_{10}$ and S$_{11}$ respectively, caused by the energy proximity and coupling among these states after vibrational relaxation.
The first two excited states are also populated but the decay is faster from S$_3$.
The diabatic transfer from 2\Ag\ (S$_5$) and 2\Bunog\ (S$_8$) are the fastest, where more than $70\%$ of the population is lost before 10 fs.

\section{Results: excited state quantum dynamics of circumcoronene and comparison to coronene  \label{sec:circum}}

If we think about coronene as the smallest portion of a graphene sheet including all the symmetry operations (\Dseish), then circumcoronene is the next step towards the generalization of graphene retaining the highest symmetry. 
Thus, we extended our study to this system, with twice as many atoms (72) as coronene and 210 vibrational modes.

\subsection{Vertical absorption and static results}
For consistency, the same methodology and level of theory introduced in Sec.~\ref{sec:compdetails}, and used for coronene in Sec.~\ref{sec:coro_abinitio}, has been used to characterize the diabatic PES of circumcoronene. 
The vertical excitations have been computed also for this bigger graphene nanoflake using TD-DFT and included in Table~\ref{tab:vert-circum}, similarly to the analysis of coronene in Table~\ref{tab:vert-excited},
except for the absence of the EOM-CCSD and experimental references. 
Although this system has been synthesized recently by R. Gershoni-Poranne \emph{et al.},~\cite{GeTs23} there is not an experimental absorption and emission spectrum, yet. 
One very interesting outcome of this work is the fact that the molecular orbitals involved in the excitations are equivalent to those in coronene, however, the excited state energies lay one electronvolt beneath due to delocalization effects.
\begin{table}[h]
    \caption{%
      Vertical excitation energies, in eV, and their corresponding oscillator strengths were obtained using TD-B3LYP-D3/def2-SVP 
      level of theory for circumcoronene in \Ddosh\ symmetry. 
      \label{tab:vert-circum}
    }
    \centering
    \begin{tabular}{@{\ } r r@{\ \ \ \ } l@{\ \ } c@{\ \ \ \ \ } r@{\ } l@{\ \ } c@{\ \ } }
      \hline\hline\\[-0.75cm]
        State   & 
        \multicolumn{2}{c}{~~~~TD-B3LYP-D3/def2-SVP} &
        & \multicolumn{2}{l}{Coronene$^{(Tab.~\ref{tab:vert-excited})}$ } \\
      \hline\\[-0.75cm]
~~~S$_0$        &       ~~~~~~~~~~~~0.00    &             &&     X(A$_{\text{g}}$)     &   (S$_0$)  \\[-0.15cm]
 S$_1$        &       2.24    &     (0.00)  &&     1\Bdosu              &   (S$_1$)  \\[-0.15cm]
 S$_2$        &       2.43    &     (0.00)  &&     1\Btresu              &    (S$_2$) \\[-0.15cm]
 S$_3$        &       3.09    &     (1.27)  &&     2\Btresu              &   (S$_7$)  \\[-0.15cm]
 S$_4$        &       3.09    &     (1.27)  &&     2\Bdosu               &   (S$_6$)  \\[-0.15cm]
 S$_5$        &       3.10    &     (0.00)  &&     ~~1\Ag                &   (S$_4$)  \\[-0.15cm]
 S$_6$        &       3.10    &     (0.00)  &&     ~~1\Bunog             &   (S$_3$)  \\[-0.15cm]
 S$_7$        &       3.13    &     (0.00)  &&     2\Bunog               &   (S$_8$)  \\[-0.15cm]
 S$_8$        &       3.19    &     (0.00)  &&     ~~2\Ag                &   (S$_5$)  \\[-0.15cm]
 S$_9$        &       3.21    &     (0.00)  &&     3\Bunog               &   (S$_9$)  \\[-0.15cm]
 S$_{10}$     &       3.31    &     (0.00)  &&     ~~3\Ag                &   (S$_{11}$)  \\[-0.15cm]
 S$_{11}$     &       3.31    &     (0.00)  &&     4\Bunog               &   (S$_{10}$)  \\[0.15cm]
      \hline\hline
    \end{tabular}
\end{table}
Table~\ref{tab:vert-circum} includes also the relation of each excited state with its counterpart in the coronene molecule. 
This analysis was done by looking at the MOs involved in every transition and comparing those orbitals with the ones involved in the coronene excitations.

\begin{figure}[h]
  \begin{center}
    \includegraphics[width=0.6\linewidth]{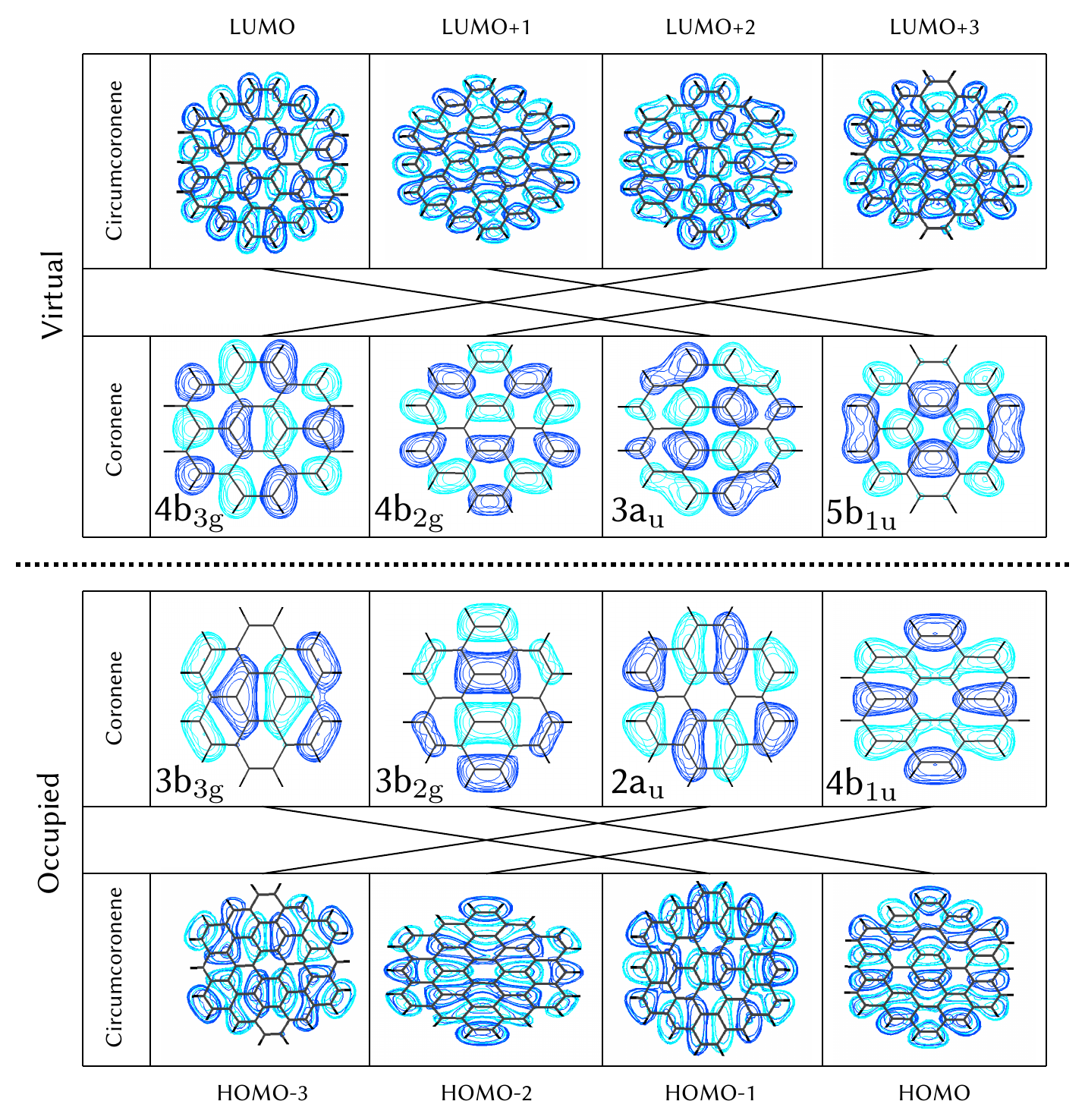}
    \end{center}     
    \caption{%
      Top view of molecular orbitals of circumcoronone showed in comparison with respect to the corresponding ones in coronene. 
      All these orbitals correspond to $\pi^*$ orbitals asymmetric with respect to the molecular plane.
      Symmetry labels are included for the coronene system, connected with lines with equivalent MO in circumcoronene, 
      where the symmetry can be extrapolated. 
      }
    \label{fig:circum_mo}
\end{figure}

The graphical representation of these MOs is included in Fig.~\ref{fig:circum_mo}, where the four highest occupied and four lowest virtual MOs of coronene and circumcoronene have been plotted. 

These MOs in circumcoronene are pairwise degenerated, similar to the coronene orbitals. The symmetry of each molecular orbital has been deluded by comparison with respect to the equivalent MO in coronene. The correlation between the MOs in the two systems has been represented by crossing lines in Fig.~\ref{fig:circum_mo}.
It is interesting that the MOs that contribute the most to the bright excited states in coronene, 2\Bdosu (S$_6$) and 2B$_{3\text{u}}$ (S$_7$), 
correspond to the same MOs that are involved in the two excited states in cirumcoronene showing a non-zero oscillator strength, 2\Bdosu\ (S$_3$) and 2\Btresu\ (S$_4$).

\newpage

\begin{figure}[h]
  \begin{center}
    \includegraphics[width=0.7\linewidth]{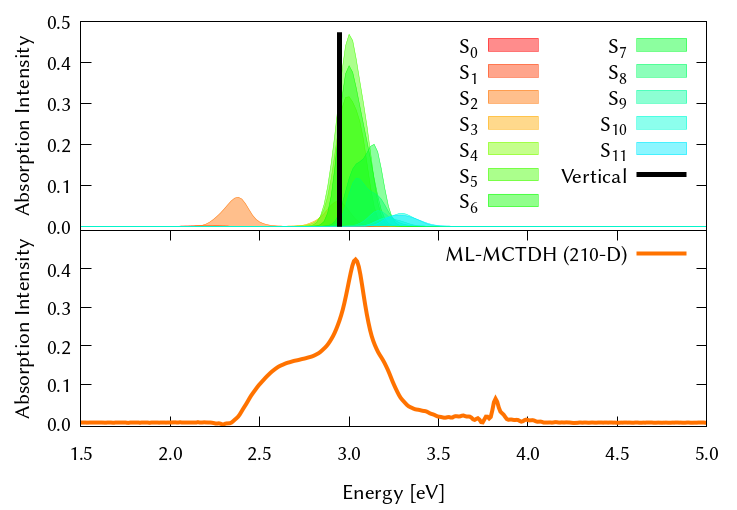}
    \end{center}     
    \caption{%
      Computed absorption spectrum of circumcoronene. 
      Top panel (a) shows the static spectra computed by vertical excitations from the ground state optimised geometry using TD-DFT (black line)
      and including vibronic effects by computing vertical excitations  
      from 200 initial conditions selected from a Wigner distribution, which are shown in colors after convoluting with gaussians with a width of 0.2 eV.
      Lower panel (b) represents the dynamic absorption spectra obtained from the Fourier Transformation (FT) of the autocorrelation function $\langle\Psi(t)|\Psi(t=0)\rangle$ of the quantum dynamics starting from 1\Btresu\ (S$_2$) excited state computed with ML-MCTDH and including the 210 normal modes.
      }
    \label{fig:circum_spec}
\end{figure}

The circumcoronene absorption spectrum, represented in Fig.~\ref{fig:circum_spec}, has been simulated in the same manner we did for coronene in Fig.~\ref{fig:coro_spec}, where the upper panel shows the static spectrum with the vertical absorption peak allowed by symmetry (S$_3$ and S$_4$) and including then vibronic effects by computing vertical transitions from 200 initial configurations generated from a Wigner distribution, which reveals a smaller absorption peak at lower energies, corresponding to excite the system to S$_2$.  

\newpage
\subsection{Excited state quantum dynamics from every electronic state in the coupled manifold\label{sec:circum_dyn}}

The same procedure has been employed to study the out-of-equilibrium dynamics of circumcoronene to observe the effect of the enlargement of the system in the molecular plane.
By exciting the system to the second singlet excited state (1\Btresu, S$_2$) the dynamic absorption spectrum has been recovered from the Fourier Transformation of the autocorrelation function and plotted in the bottom panel of Fig.~\ref{fig:circum_spec} to compare better with respect to the static absorption. The quantum dynamics for this system have been studied using ML-MCTDH on the LVC precomputed potentials calculated at the TD-B3LYP-D3/def2-SVP including all the vibrational degrees of freedom. During these dynamics, the grid sizes were 5.91$\times 10^{219}$ and 2.43$\times10^{50}$ nuclear configurations were included. Due to its large size, on-the-fly TSH dynamics is out of reach for this system. 

\begin{figure}[h]
  \begin{center}
    \includegraphics[width=0.7\linewidth]{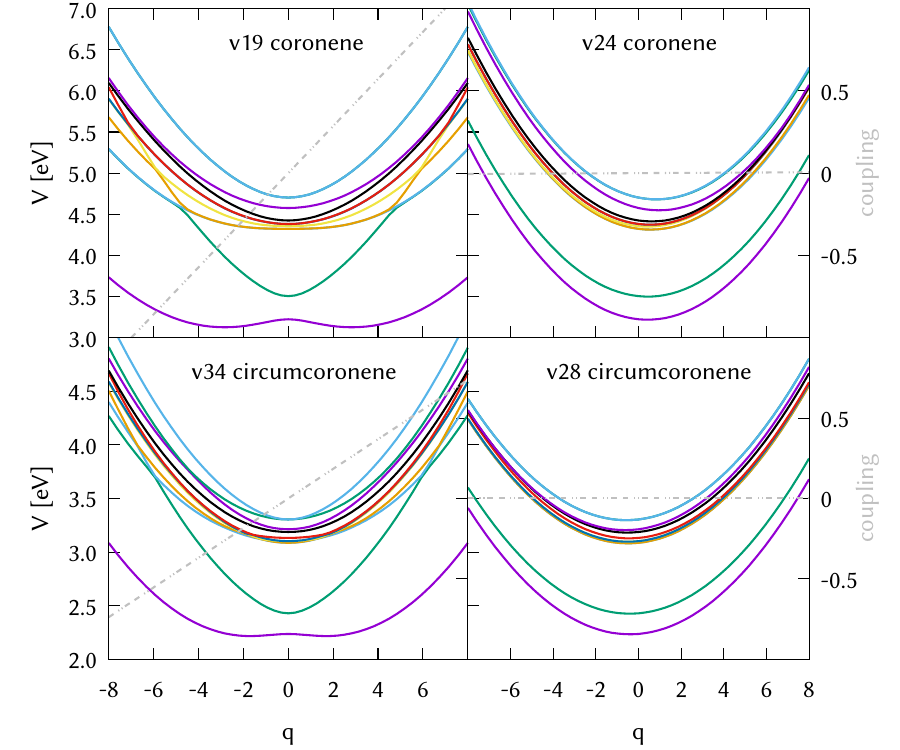}
    \end{center}     
    \caption{%
      Potential energy cuts along the two leading modes in the dynamics from the excited state 1\Btresu (S$_2$) of coronene (top panels) and the corresponding modes in circumcoronene (bottom panels). 
      The right panels correspond to the most important symmetric breathing modes with negligible coupling and large gradients, 
      while the left panels correspond to the most relevant scissoring mode in both systems, exhibiting large diabatic couplings.
      The diabatic coupling between 1\Bdosu (S$_1$) and 1\Btresu (S$_2$) is plotted in grey. 
      }
    \label{fig:pes_cuts}
\end{figure}

We also performed a vibrational mode analysis by exploring the analogous modes that were relevant in coronene. 
The scissoring mode in coronene ($\nu_{19}$) is equivalent to $\nu_{34}$ in circumcoronene, which is equally important for the de-excitation dynamics from S$_2$ in this molecule. 
In the same way, the totally symmetric breathing $\nu_{24}$ in coronene is equivalent to $\nu_{28}$ in circumcoronene, which is also responsible for leading the molecules away from the Frank-Condon region on the S$_2$ state.
In Fig.~\ref{fig:pes_cuts}, these four cuts of the adiabatic PES are represented in order to show the clear correlation between these important vibrations in both systems. 

%


%
Although the behavior of both PES is quite similar, it is important to emphasize here that they are displaced to lower energies and that there is a higher density of states in the case of circumcoronene, which can be explained by the size difference and the larger electronic delocalization in this system.

\begin{figure*}[h]
  \begin{center}
    \includegraphics[width=0.8\linewidth]{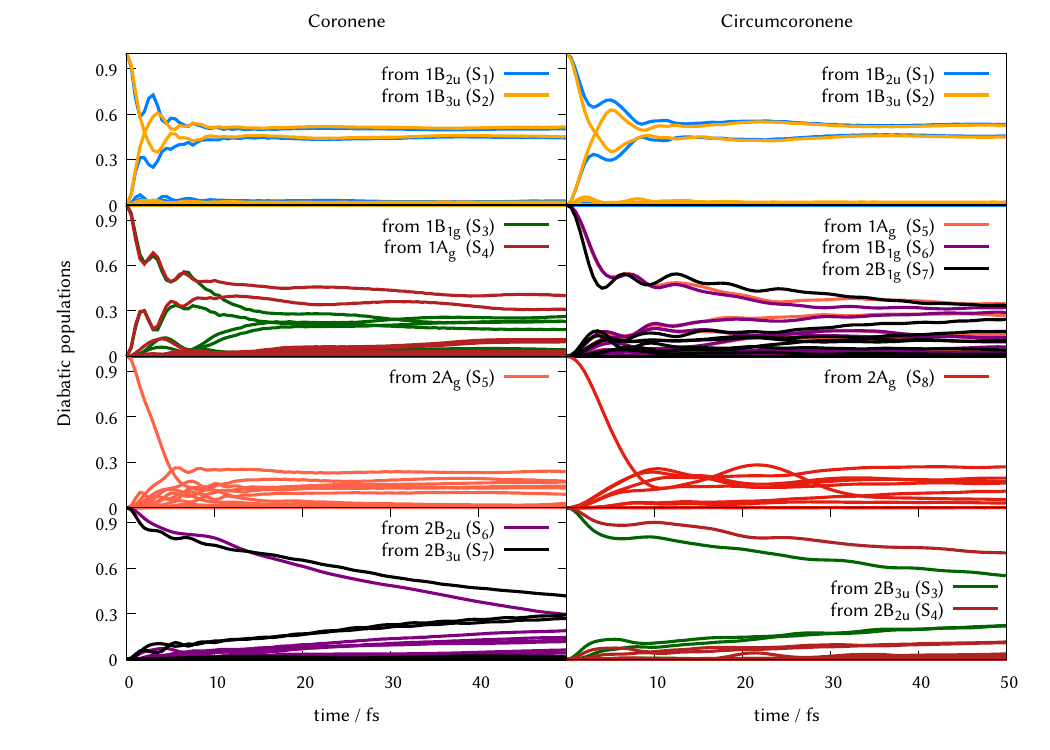}
    \end{center}     
    \caption{%
      Population transfer from ML-MCTDH full-D dynamics starting in every excited state for 
      coronene (left column) 
      and circumcoronene (right column). 
      The order of states has been altered to keep the same order as in Table~\ref{tab:vert-circum} so that the dynamics are comparable.
      }
    \label{fig:allpops}
\end{figure*}

In Figure~\ref{fig:allpops} we grouped the excited state dynamics in coronene and circumcoronene from S$_2$ to S$_8$ according to their similarities. 1\Bdosu\ and 1\Btresu\ states behave in a very similar manner in coronene and circumcoronene, transforming population among them in the sub 20-fs timescale. After this time, an equilibrium is reached and both states keep half the population. 1\Bunog, 1\Ag, and 2\Bunog\ states have an equivalent initial decay
lead by asymmetric in-plane motions. The population of the states is kept constant after 20 fs. When exciting to 2\Ag\ in coronene and circumcoronene (S$_5$ and S$_8$ respectively) a complete population decay to other states is reached within 10 fs. The bright 2\Bdosu\ and 2\Btresu\ states, however, decay at a slower pace, showing lifetimes above 250 fs.

\section{Conclusions}
\noindent
We have benchmarked the level of theory to describe the electronic potential energy surfaces in coronene, finding that TD-B3LYP-D3/def2-SVP provides a good agreement with experimental results. A parametrised LVC model explains the symmetry-forbidden vibronic band at around 3.4 eV. The full dimensional excited state dynamics is driven by breathing and tilting modes. No reduced mode combination, however, leads to the full dimensional result, indicating that most of modes contribute to the dynamics with small couplings. On-the-fly excited state dynamics where electronic quantities are computed along the trajectories show that the contribution of triplet states is negligible and indicates that the conformational molecular changes are correctly described by displaced harmonic potentials.  As the D$_{2\rm h}$ assumed symmetry mixes electronic states that would be degenerated at the FC region in the D$_{6\rm h}$ group, differences in population transfer arise in on-the-fly dynamics starting in the S$_2$ state when compared to precomputed potentials. The same LVC/TD-B3LYP-D3/def2-SVP methodology was applied for the circumcoronene molecule with 210 nuclear degrees of freedom. Although the absorption spectrum lays 1 eV below the spectrum of coronene, the spectral features are very similar. In fact, the excited state dynamics started from S$_2$ to S$_{8}$ involving twelve coupled electronic states and 210 DOFs agree very well with the coronene dynamics. This result shows that only one electronic structure calculation at the FC in the larger system and a state mapping to the smallest unit -coronene- are needed to describe the early ultrafast dynamics of polyaromatic hydrocarbons and their derivatives.

\section*{Data availability statement}

The benchmark for the vertical energies in coronene, the comparison from spectra computed from 8D dynamics and full dimensionality for coronene, the ML-MCTDH trees which show the nuclear base function distribution among the 102 and 210 degrees of freedom and the detail from the dynamics from every electronic state for both molecules can be found in the Supporting Information file. The frequency calculations at the optimised geometries of both compounds, quantics inputs, operator files and initial conditions for the on-the-fly TSH dynamics are in the comp$\textunderscore$data.zip file.

\section*{Conflicts of interest}
There are no conflicts to declare.

\section*{Acknowledgements}
SG acknowledges NextGenerationEU funds (Mar\'ia Zambrano Grant
for the attraction of international talent) and the USAL grant "Programa Propio C1". The authors thank the funding by Spanish Ministry of Science and Innovation (MCIN/AEI/10.13039/501100011033) grant No. PID2020-113147GA-I00.

\bibliography{mybib}

\end{document}